\documentclass[sigconf,authorversion,nonacm,10pt,balance=false]{acmart}

\AtBeginDocument{
  \usepackage{hyperxmp}
}
\usepackage{amsmath}

\usepackage{amssymb}
\usepackage{enumitem}
\usepackage{multirow}
\usepackage[labelfont=bf,textfont=md,skip=3pt,font=footnotesize]{caption}
\usepackage{subcaption}
\usepackage{balance}
\usepackage{adjustbox}

\newcommand{\name}{\texttt{AV-Twin}}
\newcommand{\header}[1]{\vskip 0.1cm \noindent{\bf #1}}

\usepackage[addedmarkup=uline, defaultcolor=magenta, todonotes={textsize=tiny, textwidth=1.4cm}, authormarkuptext=name, commandnameprefix=ifneeded]{changes}
\sethighlightmarkup{{\IfIsColored{\color{authorcolor}}{}#1}}
\setauthormarkup{}
\definechangesauthor[name=Zhao, color=red]{mm}

\definecolor{myblue}{rgb}{0.34, 0.54, 0.99}

\newcommand{\textblue}[1]{\textcolor{myblue}{#1}}

\definechangesauthor[name=Lan, color=green!75!blue]{zt}

\definechangesauthor[name=Tang, color=green!50!blue!75]{yw}

\definechangesauthor[name=Wang, color=green!50!blue!25]{yh}

\definechangesauthor[name=Lai, color=red!15!blue]{hw}

\begin{document}

\title{Building Audio-Visual Digital Twins with Smartphones}


\author{Zitong Lan, Yiwei Tang, Yuhan Wang, Haowen Lai, Yiduo Hao, Mingmin Zhao}
\email{{ztlan, tgg123, yyhhwang, hwlai, yiduohao, mingminz} @seas.upenn.edu}
\affiliation{%
  \institution{University of Pennsylvania}
  \country{USA}
}



\begin{abstract}

Digital twins today are almost entirely visual, overlooking acoustics—a core component of spatial realism and interaction. We introduce \name{}, the first practical system that constructs editable audio-visual digital twins using only commodity smartphones. \name{} combines mobile RIR capture and a visual-assisted acoustic field model to efficiently reconstruct room acoustics. It further recovers per-surface material properties through differentiable acoustic rendering, enabling users to modify materials, geometry, and layout while automatically updating both audio and visuals. Together, these capabilities establish a practical path toward fully modifiable audio-visual digital twins for real-world environments.
We provide a \href{https://youtu.be/k31nKDRhJJw}{\textblue{demo video}} for system.

\end{abstract}





\settopmatter{printfolios=true}


\maketitle

\vspace{-10pt}
\section{Introduction}
\label{sec:introduction}

\begin{figure}[t]
    \centering
    \includegraphics[width=0.9\linewidth]{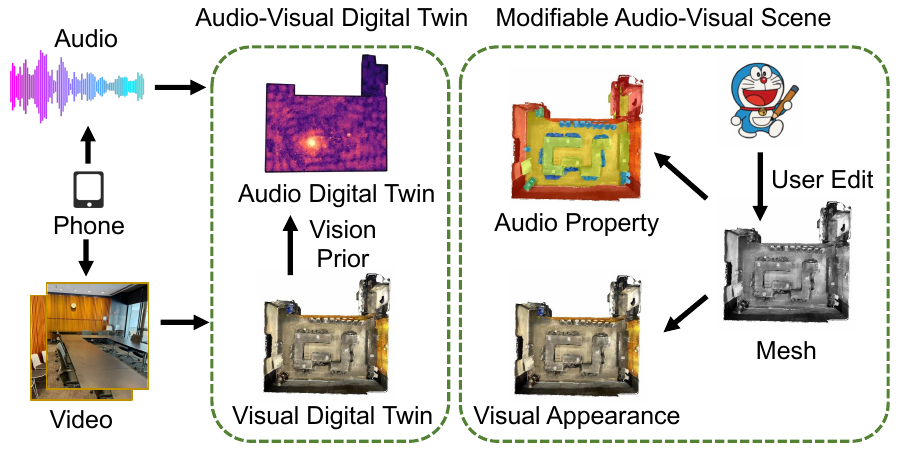}
    \caption{Users can easily reconstruct an audio-visual digital twin using only a pair of smartphones. \name{} further extends this to a modifiable audio-visual scene by estimating the material properties of each mesh and enabling both material and geometry edits.}
    \label{fig:teaser}
    \vspace{-15pt}
\end{figure}

Digital twins, computational replicas of physical environments, are rapidly emerging as a foundational technology across AR/VR, robotics, architecture, smart buildings, and human–machine interaction~\cite{iliuctua2024digital, sharma2022digital, gallala2022digital, lyu2024digital}.
They allow users to simulate how a physical environment would behave under varying conditions, evaluate designs before deployment, and build interactive experiences grounded in the real world.
The promise of digital twins hinges on two fundamental requirements: fidelity and modifiability~\cite{tao2017digital, botin2022digital}.
Fidelity ensures that the digital replica mirrors the real world with realistic and accurate behavior, while modifiability enables users to change the virtual environment and observe the resulting effects as if those modifications occurred in reality. 

Despite its broad application, today's digital twin primarily focuses on the visual modality.
Advances in computer vision and graphics can build detailed geometric models, realistic textures, and editable meshes that support a wide range of design and simulation tasks~\cite{huang2024sc, bao20253d, borycki2024gasp, chen2024lidar}.
However, a truly realistic digital twin must be audio-visual, because sound is one of the fundamental modalities that shapes how humans and intelligent systems perceive and interact with physical spaces.
For instance, AR/VR realism depends on both the acoustic effects, i.e. how sound reflects and reverberates~\cite{luo2023learningneuralacousticfields, lan2024acoustic} and the actual audio contents~\cite{lan2025guiding, tian2025audiox, hai2024ezaudio}.
auditorium and classroom design requires acoustic simulation~\cite{milo2020acoustic, manesh2024acoustic, broyles2022design};
and robots and smart devices rely on acoustic cues for navigation, localization, and sensing~\cite{chen2024sim2real, chen2020learning, fan2020acoustic}.
Without acoustics, a digital twin omits a fundamental perceptual and functional dimension of real environments.

While acoustics is essential for a multimodal digital twin, acoustic digital twins remain far less developed than their visual counterparts.
The core difficulty lies in the very nature of sound propagation: capturing how sound travels, reflects, and attenuates throughout a space -- essentially capturing the acoustic field -- requires measuring the room impulse response (RIR) across the environment.
This stands in contrast to visual geometry, which can be reconstructed from a handful of images.
Each acoustic measurement is merely an aggregation of sound arriving from all directions.
As a result, capturing an acoustic field requires densely sampled and spatially distributed measurements.
Current solutions for acoustic field reconstruction are too costly and time-consuming for practical uses~\cite{chen2024real, GTU-RIR, koyama2021meshrir}.
These systems typically require wired speaker–microphone arrays, motion-capture equipment, and motorized rails, often costing over \$100k and requiring many hours of measurement even for modest spaces~\cite{chen2024real, GTU-RIR}.
Consequently, much of the recent research on room acoustics~\cite{luo2021neural, su2022inras, liang2023neural, ratnarajah2022mesh2ir, ick2025data} relies on simulations or on a handful of pre-collected datasets that span only a few environments~\cite{liang2023av, lan2024acoustic, chen2024av, bhosale2024av}.
Beyond the difficulty in capturing acoustic fields, current acoustic field models are not modifiable.
State-of-the-art methods~\cite{lan2024acoustic, luo2022learning, su2022inras} represent acoustic field with neural implicit representations, which entangle contributions from all surfaces and objects.
As a result, changing materials, removing furniture, or testing alternative layouts is not possible without full re-capture.

In this paper, we propose \name{}, the first practical system that builds an audio-visual digital twin with high fidelity and modifiability.
As illustrated in Fig.~\ref{fig:teaser}, \name{} uses the visual (i.e., cameras) and acoustic (i.e., microphone–speaker) sensors on commodity smartphones to build an audio-visual replica of the physical world.
While building an acoustic digital twin in isolation is prohibitively measurement-heavy, our key insight is that visual cues offer information that acoustics modality alone cannot obtain efficiently.
Specifically, visual observations reveal scene layout that eliminates much of the ambiguity in acoustic reconstruction.
Moreover, when it comes to modification, \name{} leverages visual cues to construct a mesh-based audio-visual scene graph, where each primitive carries both visual appearance and acoustic properties.
This visually derived structure allows users to edit or animate the scene (e.g., changing materials, removing furniture, or altering layout) just as they would in a visual digital twin.
\name{} then propagates these edits through both modalities, updating the visual rendering and recomputing the corresponding acoustic behavior.

Delivering \name{}'s audio-visual digital twin requires a series of key innovations, which we described below.

\header{Practical and Efficient Acoustic Field Capture.}
At the core of acoustic field reconstruction is to capture room impulse response (RIR), which records how an emitted acoustic pulse travels through the environment, including all reflections and multipath components. 
\name{} achieves practical and efficient acoustic field capture with the following designs, each addressing a major bottleneck in prior workflows.
\emph{(1) Smartphone-only RIR capture.}
To eliminate dedicated hardware, \name{} enables full RIR capture using only commodity smartphones. Two users can walk naturally through the space while their phones emit probe signals and record the resulting audio.
To support wireless measurement, \name{} designs an acoustic protocol with chirp signal to synchronize between phones while measuring the acoustic RIRs.
With the fine sampling resolution of audio hardware (e.g., 48 kHz) and direct access to raw samples through mobile OS, this untethered design achieves sub-ms synchronization.
\emph{(2) Visual digital twin for RIR spatial grounding.}
RIRs are only meaningful when tied to device locations.
While prior systems use a motion-capture system~\cite{chen2024real}, \name{} leverages the visual digital twin simultaneously constructed via mobile SLAM.
This visual SLAM provides globally consistent device trajectories in the reconstructed 3D scene, allowing every RIR to be spatially grounded.
\emph{(3) Dynamic-trajectory RIR collection.}
Conventional approaches measure RIRs exhaustively across a dense Tx–Rx grid: with N transmitters and M receivers, they must collect N×M RIRs. Such exhaustive sampling can take more than 50 hours for 2,000 RIRs ~\cite{GTU-RIR}.
\name{} replaces this rigid grid with a dynamic-trajectory capture paradigm, where users simply walk through the environment while their smartphones are continuously recording RIRs.
This natural motion samples a wide variety of spatial locations and exposes diverse multipath propagation.
In practice, a short 20 mins walkthrough
suffices to recover the acoustic field, over 100× more efficient than a grid-based approach.
\emph{(4) Visual-assisted acoustic field modeling.}
To improve efficiency, \name{} introduces visual-assisted acoustic volume rendering (AVR) to model the acoustic field with structural guidance provided by the visual digital twin.
During both training and inference, visual-assisted AVR casts rays from the microphone to the mesh and predicts the re-transmitted signal only at the first ray-surface hit point. 
Consequently, our method evaluates only one physically valid surface hit per ray, instead of exhaustively sampling $N$ points along each ray as done in prior AVR~\cite{lan2024acoustic}.
It then aggregates these contributions with physically grounded time delays and amplitude decay. 
This method achieves 10× faster rendering speed and 2× higher data efficiency than vanilla AVR.
When combining all these design, we achieve an efficient mobile acoustic field capture pipeline to build the audio-visual digital twin.

\header{Modifiable Audio-Visual Scene.}
Now that we have described how \name{} efficiently builds the acoustic field on a mobile device, the remaining challenge is that this acoustic digital twin is not modifiable.
This is because state-of-the-art neural acoustic field models represent room acoustics as implicit, entangled functions over the entire space. 
While these models achieve high fidelity, they offer no mechanism for editing: users cannot remove a wall, change a material, or test an acoustic treatment.
Our key innovation is to transform this implicit acoustic field into an explicit, object-aware audio-visual scene graph.
Leveraging geometry and object boundaries obtained from the visual modality, \name{} (i) disaggregates multipath energy into per-mesh acoustic contributions, and (ii) estimates the acoustic properties of each surface, such as reflectivity.
However, recovering material properties is inherently challenging: each RIR is a superposition of many acoustic paths and no single waveform directly exposes the properties of an individual surface.
\name{} addresses this challenge by combining multiple RIRs recorded across the scene with differentiable acoustic rendering to estimate the material properties.
To make this problem tractable, we incorporate vision priors. 
Visually similar and spatially adjacent surfaces (e.g., walls, door, blackboards) tend to share similar material properties. 
Grouping them together reduces the dimensionality of the estimation problem and stabilizes learning.
Material reflectivities and device patterns are treated as parameters, which are optimized so that the synthesized RIRs closely match the measured ones. 
Through this optimization, the complex multipath effects can be factored back into per-material properties.
This produces a representation in which visuals and acoustics are tied to the same set of scene primitives: meshes with visual appearance and acoustic properties.
This explicit representation enables modifiability.
Users can change materials, remove or insert furniture, adjust room layouts, or animate objects, and \name{} automatically updates the corresponding audio-visual observations.
In effect, \name{} brings to acoustics what mesh-based scene graphs brought to vision: a foundation for editing and simulation within a digital twin.

We build a holistic system solution to capture the audio-visual digital twin and make it editable.
\name{} builds an iOS App runs in real time for users to efficiently capture RIRs; our proposed visual-assisted AVR model reconstructs the acoustic field from the captured RIRs; and our material parameter estimation method further enables scene editing and modifications in the digital twin.
We extensively benchmark our captured AV-Twin for each sub-module: (1) RIR capture accuracy. (2) Performance of acoustic field reconstruction. (3) Material property estimation accuracy for scene editing. 
Our mobile RIR capture component achieves an average ToF estimation error of 0.1~ms and a detection rate of 99.6\%. 
Our dynamic-trajectory-based method shows more than 100× improvement in data collection efficiency compared to traditional methods.
Our visual-assisted AVR can further improve the data efficiency by 2x and improve the acoustic field rendering speed by 10x.
For acoustic property estimation, we achieve a mean absolute error of 5.6\% in reflection coefficients estimation and a correlation coefficient of 0.96 to the fixed measurement setup. 
A user study shows that 88\% of participants preferred our dynamic-trajectory-based method over conventional setups. 
Another user study shows that over 90\% of users think the editing in the visual scene also matches with the audio scene.
We also show that augmenting a localization model with acoustic field model reduces error by 50\% and achieves sub-meter accuracy (0.45 m).

The key contributions of the paper are as follows:
\begin{itemize}[leftmargin=12pt,itemsep=0pt,topsep=-4pt]
\item We introduce \name{}, the first system that constructs \emph{audio-visual digital twins} with commodity smartphones.
\item We present a practical acoustic field reconstruction framework with smartphones to collect RIRs grounded with a visual digital twin. Our dynamic trajectory method achieves much less measurement time.
\item We design a visual-assisted AVR that leverages the room geometry from a visual digital twin to improve both data efficiency and rendering speed.
\item We introduce a vision-guided differentiable material-property estimator that recovers per-surface reflectivities for modifiable digital twins.
\end{itemize}

\section{Related Work}
\label{sec:related-work}

\header{Acoustic field modeling.}
Capturing acoustic fields in real-world environments is crucial for studying sound propagation and building models for immersive audio. 
However, real-world acoustic capture is both time-consuming and resource-intensive~\cite{chen2024real, GTU-RIR, koyama2021meshrir, liang2023av}. 
RAF~\cite{chen2024real} requires specialized rigs costing over \$100k, and GTU-RIR~\cite{GTU-RIR} reports that collecting just 400 RIR samples with a single microphone can take 10 hours.
MeshRIR~\cite{koyama2021meshrir} involves bulky setup and is hard to reposition to other scenes. 
Building upon these datasets, ML research models the acoustic field as continuous functions~\cite{luo2022learning, su2022inras, lan2024acoustic, liang2023av, liang2023neural, bhosale2024av, chen2024av, lan2025resounding}. 
They use neural implicit representations~\cite{luo2021neural, su2022inras, chen2024av} to model RIR directly from arbitrary speaker-microphone locations.
It is further advanced by acoustic volume rendering that encourages multi-view consistency~\cite{lan2024acoustic}.
\name{} enables efficient acoustic field reconstruction compatible with those methods. 

\header{Acoustic localization.}
Prior acoustic localization methods usually rely on multiple speakers and microphones. Many use frequency-modulated continuous waves to estimate ToF and trilaterate positions with multiple devices~\cite{AAMouse, CAT, wang2022faceori, wang2019millisonic, gao2022mom, ge2020acoustic, zhang2017soundtrak, wu2024enabling}. Single-anchor approaches~\cite{AcouRadar, SA2, cai2024locating} reduce device requirements by modeling frequency- or angle- dependent responses with one speaker, but require extensive calibration and struggle in multipath-rich rooms. More recent methods design 3D-printed metasurfaces~\cite{owlet, spidr} to create location-dependent acoustic signatures, while echo-based techniques~\cite{tung2015echotag, SyncEcho, onemic} rely on dense wall-reflection fingerprints and fixed sensor orientation.

\header{Acoustic sensing}.
Acoustic sensing has emerged as a powerful approach in mobile and ubiquitous computing.
It has been widely studied for ranging and localization~\cite{peng2007beepbeep, lazik2015ultrasonic}, vital sign detection~\cite{liu2022enabling, song2020spirosonic, wan2023multi, wang2018c, su2024embracing, su2025manipulation, zhang2023vecare, wang2022loear}, gesture recognition, and hand-motion tracking~\cite{amesaka2022gesture, cao2023powerphone, li2022room, cao2020earphonetrack, wang2016device, wang2025ram}, and even for applications such as hearing screening~\cite{chan2023wireless}. To further push performance, researchers have introduced hardware aids such as metasurfaces to boost acoustic range and robustness~\cite{owlet, zhang2023acoustic, mallejac2025active, he2024cw}. Collectively, these efforts highlight the potential of commodity microphones and speakers as versatile sensors.
Most of this work, however, leverages acoustics to sense human or device activities at close range. 
Our focus is on building audio-visual digital twin and make it modifiable.

\section{Overview}
\label{sec:overview}

\begin{figure}[h]
    \centering
    \vspace{-10pt}
    \includegraphics[width=0.75\linewidth]{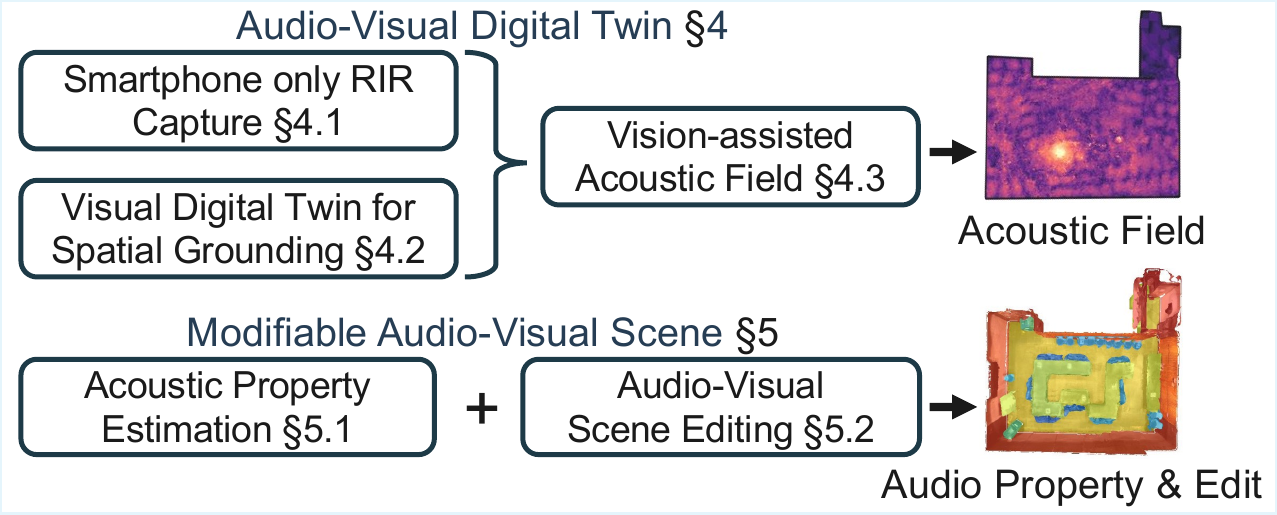}
    \caption{\name{} builds audio-visual digital twin (\S\ref{sec:audio_visual_digital_twin}) with a series of key innovations.
    It also enable modifiable audio-visual scene by estimating acoustic property (\S\ref{sec:acoustic_property_capture}) to support various editing capabilities (\S\ref{sec:audio_visual_scene_editing}).
    They enable practical applications demonstrated in the experiments.}
    \vspace{-18pt}
    \label{fig:overview}
\end{figure}

\name{} constructs a unified \emph{audio–visual digital twin} of indoor environments.
As shown in Fig.~\ref{fig:overview}, to construct an audio-visual digital twin, \name{} collects RIRs with a pair of smartphones with dynamic trajectories (\S\ref{sec:acoustic_protocol}). 
These RIRs are spatially grounded by a complementary visual digital twin (\S\ref{sec:slam}).
We also introduce a visual-assisted acoustic field model (\S\ref{sec:acoustic_field_capture}) to reconstruct the acoustic field efficiently.
Beyond these, \name{} extends to a modifiable audio-visual digital twin by estimating the acoustic properties and assigning to each mesh in the visual digital twin (\S\ref{sec:acoustic_property_capture}).
We employ a differentiable rendering model to recover the material properties.
This enables various audio-visual scene editing (\S~\ref{sec:audio_visual_scene_editing}).
\name{} supports downstream applications including immersive audio rendering, interactive scene editing, and acoustic localization (\S~\ref{subsec:evaluation_applications}).

\section{Audio-visual digital twin}
\label{sec:audio_visual_digital_twin}

Building a complete audio–visual digital twin requires not only capturing how a room looks, but also how it sounds.
To achieve this, we first introduce a mobile RIR capture system using a single pair of smartphones with dynamic trajectories (\S\ref{sec:acoustic_protocol}).
We then spatially anchor the captured RIRs using the visual digital twin reconstructed from the same smartphones (\S\ref{sec:slam}).
Finally, we introduce a method to reconstruct the acoustic field with visual-assisted acoustic volume rendering with vision priors to improve efficiency (\S\ref{sec:acoustic_field_capture}).


\subsection{Smartphone-only RIR capture}
\label{sec:acoustic_protocol}

Our mobile RIR capture system can measure RIRs with just a pair of commodity smartphones.
To support this, \name{} designs an acoustic protocol to measure RIRs and synchronize between two devices.

\begin{figure}[t]
    \centering
    \includegraphics[width=0.75\linewidth]{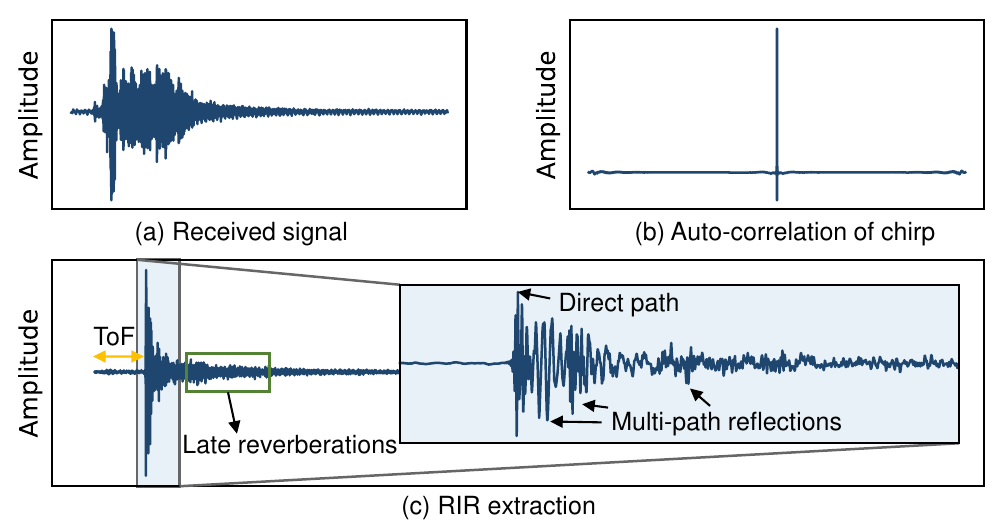}
    \caption{Illustration of RIR extraction. (a) The received signal is cross-correlated with the reference chirp. Since (b) chirp's auto-correlation produces a delta-like peak, (c) the resulting output reveals the RIR, which consists of direct path, multipath reflections and late reverberations.}
    \label{fig:rir_extraction}
    \vspace{-10pt}
\end{figure}

\header{RIR Measurement.}
An RIR $h(t)$ is a time-domain signal that characterizes how an acoustic channel responds to an impulse. It describes how sound energy emitted by a source arrives at a receiver over time, including the direct path as well as reflections and reverberations from surrounding surfaces. When the transmitter (speaker) emits a probe chirp $c(t)$, the receiver (microphone) records a signal $x(t) = c(t) \ast h(t)$.
To estimate the RIR $h(t)$, we cross-correlate the received signal $x(t)$, shown in Fig.~\ref{fig:rir_extraction}(a), with the known probe $c(t)$. 
Since $c(t)$ has a sharply peaked auto-correlation that approximates a delta function (Fig.~\ref{fig:rir_extraction}(b)), this operations reveals the RIR:
\begin{equation}
x(t) \ast c(t) = h(t) \ast (c(t) \ast c(t)) \approx h(t) \ast \delta(t) = h(t).
\label{eq:correlation}
\end{equation}
The recovered RIR (Fig.~\ref{fig:rir_extraction}(c)) contains a sharp initial peak shifted by ToF, followed by multipath reflections and late reverberations.
However, Eq.~\ref{eq:correlation} assumes that the start time of the probe signal is known. 
In practice, without precise time synchronization between Tx and Rx, the recovered RIR is shifted by an unknown offset.
We address this with the following acoustic protocol.

\begin{figure}[t]
    \centering
    \includegraphics[width=0.8\linewidth]{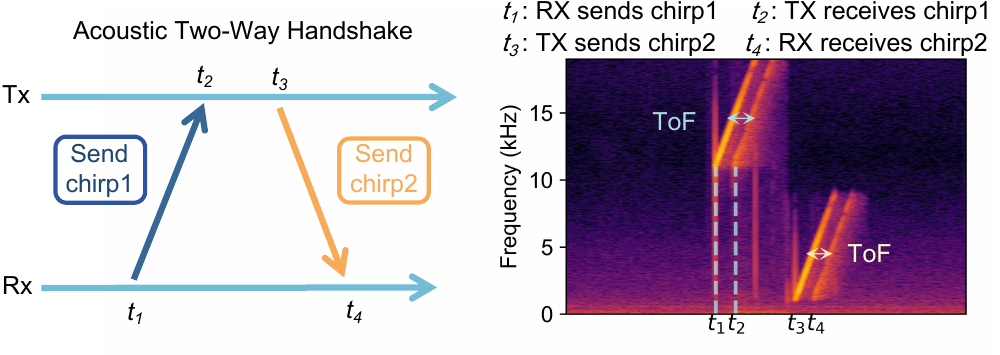}
    \caption{Illustration of acoustic two-way handshake design to simultaneously record the RIR and determine the correct ToF.}
    \label{fig:acoustic_handshake}
    \vspace{-15pt}
\end{figure}

\header{Acoustic Protocol.}
Mobile devices expose raw audio samples at high sampling rates (e.g., 48 kHz),  which makes it possible to perform synchronization directly in the acoustic domain, embedding ToF estimation into the same chirp signals used for RIR measurement.
With a 48~kHz sampling rate, each sample corresponds to $21~\mu$s, which represents the theoretical resolution limit for ToF estimation.
We use an acoustic protocol that embeds synchronization directly into the probe signals, enabling simultaneous RIR extraction and precise ToF estimation.
While prior systems use acoustic handshakes solely for ranging~\cite{peng2007beepbeep, lazik2015ultrasonic}, we reuses similar probe signals to capture full RIRs with accurate ToF.
As illustrated in Fig.~\ref{fig:acoustic_handshake}, both Tx and Rx continuously record audio. 
At $t_{1}$, the Rx emits a chirp $c_1$, which arrives at Tx by $t_{2}$. Upon detection, Tx immediately responds with chirp $c_2$ by $t_{3}$, which propagates back to Rx by $t_{4}$.
We combine the forward and backward delays to cancel the clock offset and calculate ToF as
$\frac{(t_4 - t_1) - (t_3 - t_2)}{2}$.
This formula mirrors the principle in network time protocols~\cite{peterson2007computer}, but here it is used to recover the acoustic ToF for RIR measurement.

\begin{figure}[t]
    \centering
    \includegraphics[width=0.8\linewidth]{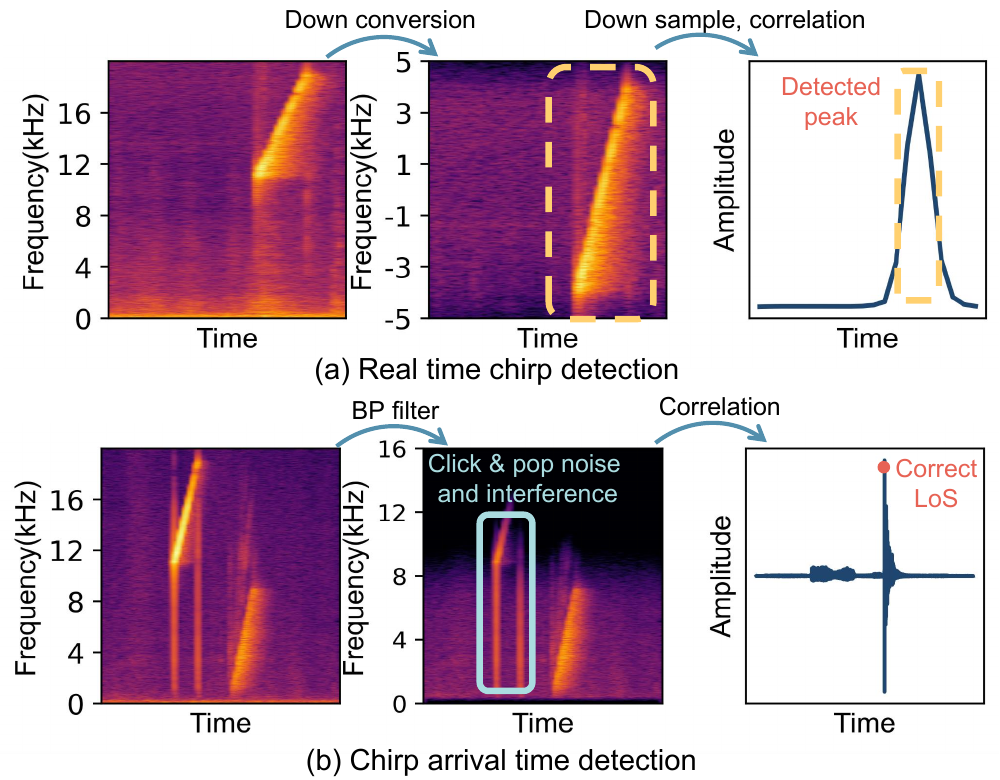}
    \caption{
(a) Real-time chirp detection: Tx recording $x_\text{tx}$ is converted to baseband and down-sampled to accelerate calculation.
It is then correlated with $c_1$ in the time-frequency domain for detection.
(b) Chirp arrival time detection: 
Correlate the recording  with the known $c_2$ in the time domain reveals the RIR and we identify the LOS peak (right).}
\label{fig:signal_detection}
\vspace{-20pt}
\end{figure}



\header{Real-time on device chirp detection.}
Fast detection of $c_1$ is critical to our design. 
If the detection latency is high, the response $t_3$ will be significantly delayed relative to the arrival time $t_2$. 
The long gap will accumulate user movement and distort the RIR estimation due to location changes.
As a result, real-time detection is necessary to mobile RIR capture.
To achieve this, we detect $c_1$ in time-frequency (TF) domain that is efficient and robust to noise~\cite{luo2021neural}.
We use the correlation coefficient between the received signal $x_\text{tx}$ and $c_1$ for detection.
This correlation coefficient is invariant to distance-dependent attenuation and peaks only when the signal $x_\text{tx}$ truly matches $c_1$.
This method supports streaming detection to improve efficiency, where only the appended audio are transformed with Fast Fourier Transform.
We also down-sample the signal to further improve efficiency.
These together reduce the detection latency from $1.1$~s to $0.1$~s.
The detection process is illustrated in Fig.~\ref{fig:signal_detection}(a).

\header{Chirp arrival time detection.}
We need to determine the chirp arrival time accurately ($t_2$ and $t_4$) at direct path, rather than at multipath or interference (middle of Fig.~\ref{fig:signal_detection}(b)).
To address this, we develop a robust detection method and demonstrate it with detection of $t_4$.
We first cross-correlate $x_\text{rx}(t)$ with $c_2(t)$, producing output $h(t)$.
We then find all candidate peaks above a threshold. 
These peaks mark the potential regions of the direct path.
We then scan these candidate peaks in ascending order.
For each candidate, we exam whether the next peak candidate is much higher than the previous candidate.
This tests whether there is a sharp increase in $h(t)$ that marks the direct path of RIR. 
Once a candidate peak passes the test, the arrival time $t_4$ is determined by that peak.
As shown in Fig.~\ref{fig:signal_detection}(b), this method is robust to interference or the false peak caused by multipath.
\header{Dynamic Trajectory Capture}
We further replace traditional grid-based sampling with a dynamic trajectory–based capture paradigm, in which users naturally walk through the environment while a smartphone continuously records RIRs. This motion densely samples spatial locations and reveals rich multipath propagation. In practice, a 20-minute walkthrough is sufficient to recover the global acoustic field, achieving over 100× higher sampling efficiency.

\vspace{-5pt}
\subsection{Visual Twin for RIR Spatial Grounding}
\label{sec:slam}

While the RIR encodes rich information about acoustic propagation, it is insufficient alone to build a complete acoustic representation. 
They are only meaningful when spatially grounded.
Therefore, we build a visual digital twin that contains both device trajectories and scene structure to spatially ground the measured RIRs.
To this end, we integrate SLAM algorithm using camera and LiDAR on the smartphone and get rid of the expensive hardware setup.

\header{SLAM for localization and scene reconstruction.}
We integrate a lightweight vision SLAM algorithm, RTAB-Map~\cite{labbe2022multi}, into our mobile platform and deploy it on both the Tx and Rx devices. 
The SLAM pipeline uses camera and LiDAR to estimate device trajectories, yielding the speaker and microphone positions ($p_\text{tx}$, $p_\text{rx}$) and orientations ($\omega_\text{tx}$, $\omega_\text{rx}$).
In addition to device poses, the SLAM output also provides room geometry $G$, represented as a mesh.


\header{Handling human interference.} 
Since our system involves  users freely scanning the scene with handheld devices, human bodies are frequently captured in the camera images and LiDAR scans, which corrupts the reconstructed scene.
To mitigate this, we record all sensor streams and perform post-processing with a YOLO segmentation model~\cite{wang2024yolov10}.
Human regions are masked out in RGB image.
To also assess whether user's body will affects RIR capturing when holding the smartphones, we also compare handheld measurements against tripod measurements and we show that it only results in minimal influence in the experiment session (\S\ref{subsec:microbenchmarks}).


\header{Aligning Tx/Rx coordinates.} 
RIR measurement requires that Tx/Rx devices share the same coordinate.
However, they have individual SLAM scanning, and their coordinates are not aligned.
To align their coordinates, we reload both databases of Tx and Rx from RTabmap and merge them to build a unified pose graph via global loop-closure detection. 
The combined graph is optimized by the general graph optimization to jointly refine all poses and produce Tx and Rx coordinates that align with each other. 
\vspace{-10pt}
\subsection{Visual-assisted Acoustic Field Modeling}
\label{sec:acoustic_field_capture}

We first introduce the formulation of acoustic field reconstruction and the limitations of prior methods.
We then introduce visual-assisted AVR to reconstruct the acoustic field efficiently with the help of a visual digital twin.

\begin{figure}[t]
    \vspace{-10pt}
    \centering
    \includegraphics[width=0.75\linewidth]{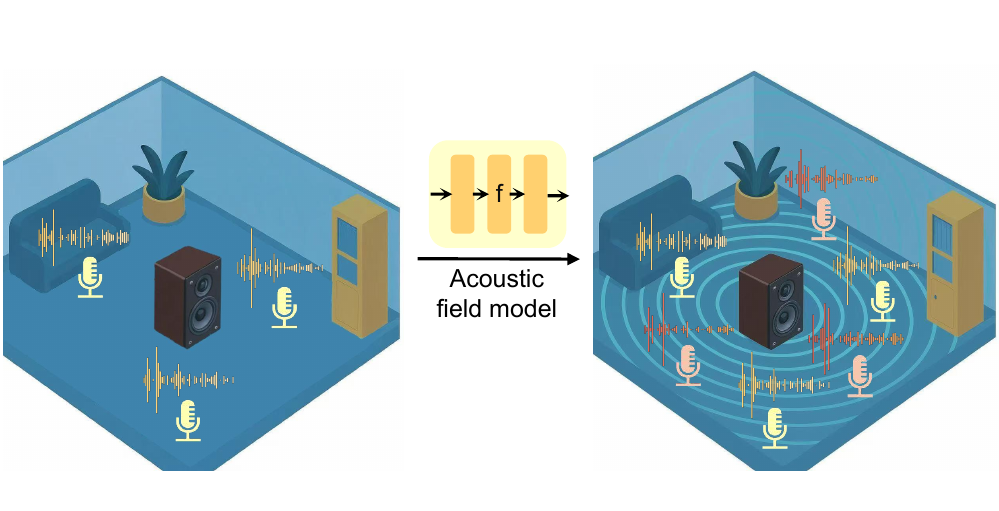}
    \caption{Acoustic field model. From limited RIR measurements, acoustic field model can understand the acoustic field propagation in the environment and can synthesize arbitrary RIRs at any microphone (and speaker) locations in the scene.}
    \label{fig:acoustic_field_model}
    \vspace{-10pt}
\end{figure}

\header{Acoustic field reconstruction.} We formulate acoustic field reconstruction as the problem of modeling a continuous mapping from speaker/microphone location to the corresponding RIR in a scene with geometry $G$.
Let $p_\text{tx}, p_\text{rx}$ denote the 3D positions of the speaker and microphone, $\omega_\text{tx}, \omega_\text{rx}$ denote their orientations.
The goal is to learn a mapping:
$f:(p_\text{tx}, p_\text{rx}, \omega_\text{tx}, \omega_\text{rx}, G) \to h(t),$
where $h(t)$ is the RIR between the speaker and microphone. 
Once the acoustic field is trained, it can generalize to any speaker/microphone locations, enabling the synthesis of RIR for arbitrary placement of speaker/microphone within the scene, as shown in Fig.~\ref{fig:acoustic_field_model}(a).

RIRs have fine-grained geometric and material dependencies.
Recent approaches model these dependencies using implicit neural representations of acoustic fields. Neural Acoustic Fields (NAF)~\cite{luo2021neural} learn local geometric features on a 3D grid and use an MLP to predict RIR spectrograms, which are transformed to the time domain. Although efficient, NAF lacks physical priors and struggles to generate high-fidelity RIRs. In contrast, Acoustic Volume Rendering (AVR)~\cite{lan2024acoustic} incorporates wave propagation principles by encoding density and re-emitted signals at scene points and aggregating contributions along rays cast from the microphone. While AVR improves fidelity, it is computationally expensive due to exhaustive pointwise network evaluations.


\begin{figure}[t]
    \vspace{-5pt}
    \centering
    \includegraphics[width=0.75\linewidth]{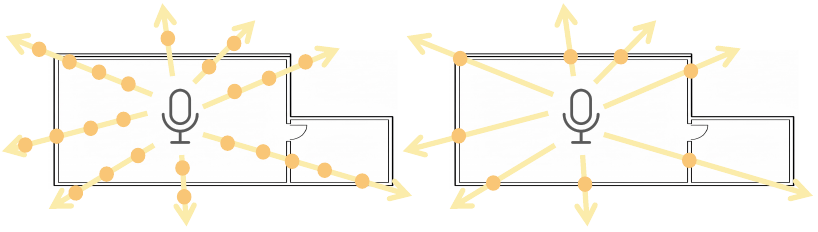}
    \caption{While AVR (left) exhaustively samples points along each ray, visual-assisted AVR (right) only sample points on the mesh surface.}
    \label{fig:avr_sampling}
    \vspace{-13pt}
\end{figure}
\header{Visual-assisted AVR.} To render high-fidelity RIRs efficiently, we propose visual-assisted acoustic volume rendering (visual-assisted AVR).
In \name{}, we recover a mesh $G$ from visual digital twin (\S\ref{sec:slam}).
Visual-assisted AVR exploits this mesh by shooting rays from the microphone and only evaluating the neural network at the first surface hit point, as shown in Fig.~\ref{fig:acoustic_field_model}(b).
This design leverages a physical prior that only surfaces can re-transmit acoustic energy, whereas empty space contributes no acoustic energy towards the RIR.
By constraining neural field queries only at these mesh surfaces, it reduces computation from sampling $N$ points exhaustively along a ray to a single point, which avoids wasting computation on vast regions of empty space.
It also helps to focus learning only on relevant surfaces, which is also data-efficient for acoustic field reconstruction.
Fig.~\ref{fig:avr_sampling} illustrates the sampling strategies in AVR and visual-assisted AVR.

Formally, given a microphone at ${p}_{\mathrm{rx}}$ and a speaker at ${p}_{\mathrm{tx}}$, we sample directions 
$\{{\omega}_k\}_{k=1}^{K}$ on the unit sphere around the microphone and cast rays:
$
{r}_k(s) = {p}_{\mathrm{rx}} + l\cdot{\omega}_k,\; l>0.
$
Each ray yields the first mesh intersection ${x}_k$.  
Unlike AVR, which samples a continuous set of volumetric points along each ray, visual-assisted AVR models each hit point $\mathbf{x}_k$ on the surface as a secondary emitter that re-transmits acoustic energy toward the microphone direction.
For each hit point ${x}_k$, the neural field predicts a re-transmitted signal:
\begin{equation}
    s_{k}(t)
= s\!\left(t;{x}_k,\omega_k,{p}_{\mathrm{tx}}, \omega_\mathrm{tx}\right),
\end{equation}
representing the acoustic signal re-emitted from that surface location toward the microphone direction $-\omega_k$.  
This network also implicitly encodes all effects related to the speaker, including position, orientation, and gain patterns.

\textit{Visual-assisted AVR rendering.}
The predicted RIR is the sum of the re-transmitted signals from all surface-hit paths:
\begin{equation}
\label{eq:avrpp_time}
h(t;{p}_{\mathrm{tx}}, \omega_{\mathrm{tx}},{p}_{\mathrm{rx}}, \omega_\mathrm{rx})
=
\sum_{k=1}^{K}
G({\omega}_k;{\omega}_{\mathrm{rx}})
\;\frac{1}{tc}\;
s_{k}\!\Bigl(t - \tfrac{d_k}{c}\Bigr).
\end{equation}
In this equation, the propagation to the microphone is modeled using time delay and free-field spherical spreading. 
Time delay is modeled by $t - d_k/c$, where $d_k = \|{x}_k - {p}_{\mathrm{rx}}\|_2$ is the distance between microphone and surface point and $c$ is the speed of sound. 
Free space signal attenuation is modeled by $\frac{1}{tc}$.
Directional gain pattern of the microphone is modeled by $G({\omega}_k;{\omega}_{\mathrm{rx}})$.
We use the same training object in \cite{lan2024acoustic} to train visual-assisted AVR.
To trace the point along the surface, we use ray casting algorithm from Open3D to trace the intersection of rays to the mesh.

\section{Modifiable audio-visual scene}
A compelling opportunity for audio–visual digital twins is the ability to move beyond reconstruction toward \emph{modifiable} audio-visual scenes.
Though acoustic field enables prediction of RIRs, it can not support modifications.
They implicitly encode wave propagation into a black-box representation, which cannot be decomposed or manipulated after training.
Unlocking truly modifiable audio–visual scenes therefore requires a different perspective: instead of representing sound propagation implicitly, we must recover the \emph{explicit physical factors} that determine how sound interacts with the environment.
Physical acoustic properties like material reflectivity provides editable, interpretable parameters that directly control the behavior of reflected and absorbed sound.  
By estimating these properties and anchoring them to the visual digital twin, a scene is decomposed into components that can be manipulated.  
This motivates our differentiable acoustic rendering framework (\S\ref{sec:acoustic_property_capture}), which infers per-surface material parameters from measured RIRs.  
Once these physical parameters are available, users can perform various audio-visual scene edits (\S\ref{sec:audio_visual_scene_editing}).

\subsection{Acoustic Property Estimation}
\label{sec:acoustic_property_capture}
RIR measures the global acoustic response of a scene and thus cannot be directly attributed to any individual single object in the environment.
To estimate these properties, we factorize the RIR into contributions from distinct acoustic paths and associate reflection parameters with each path.
Under this framework, we introduce differentiable acoustic ray tracing to estimate these parameters.

\header{Material reflectivity.}
When a sound wave encounters a surface, part of the sound wave is specularly reflected, while the rest is absorbed. 
We parameterize this with a reflection coefficient $R_s(\nu)$, defined as the ratio of outgoing to incoming amplitudes:$\frac{a_{\text{out}}}{a_{\text{in}}}\!=\!R_s(\nu)$ at the frequency of $\nu$.
These $R_s$ values are the first set of learnable parameters in our optimization.

\header{RIR in a single path.} 
For simplicity, we first assume the propagation between a speaker at $p_\text{tx}$ and a microphone located at $p_\text{rx}$ through a single path $\mathcal{P}_n$.
Along path $\mathcal{P}_n$, the acoustic wave will encounter many surface interactions that introduce attenuation.
$\mathcal{P}_n$ is defined as a sequence of points: $\mathcal{P}_n = \{ p_{n,0}\!=\!p_\text{tx}, p_{n,1}, p_{n,2}, \ldots, p_{n,K}, p_{n,K+1}\!=\!p_\text{rx} \}$.
We then define impulse response $h(t; \mathcal{P}_n, \omega_\text{tx}, \omega_\text{rx})$ for this single path, which characterizes the sound received at $p_\text{rx}$ when the speaker at $p_\text{tx}$ sends out an ideal pulse, with $\omega_\text{tx}$ and $\omega_\text{rx}$ being the orientations of speaker and microphone.
The response is influenced by the gain patterns of the speaker/microphone as well as the reflection coefficients along the path:
\begin{equation}
h_\Theta(t;\!\mathcal{P}_n) = G_\text{tx}(\omega_{n,0}; \omega_\text{tx}) \, \Gamma(t; \mathcal{P}_n, \{R_s\}) \, G_\text{rx}(\omega_{n, K}; \omega_\text{rx}),
\end{equation}
where $G_\text{tx}$ and $G_\text{rx}$ represents the learnable gain patterns of the speaker/microphone.
$\Theta$ denotes all the learnable parameters: $\Theta = \{G_\text{rx}, G_\text{rx}, \{R_s\}\}$.
$\omega_{n,0}$ and $\omega_{n, K}$ represents the outgoing ray direction from Tx position $p_\text{tx}$ to $p_\text{n,1}$ and the incoming ray direction from $p_{n,K}$ to Rx position $p_\text{rx}$, respectively.
$\Gamma(t; \mathcal{P}_n, \{R_s\})$ denotes the path impact function for the path $\mathcal{P}_n$, as described below.

\begin{figure}[t]
    \centering
    \includegraphics[width=0.8\linewidth]{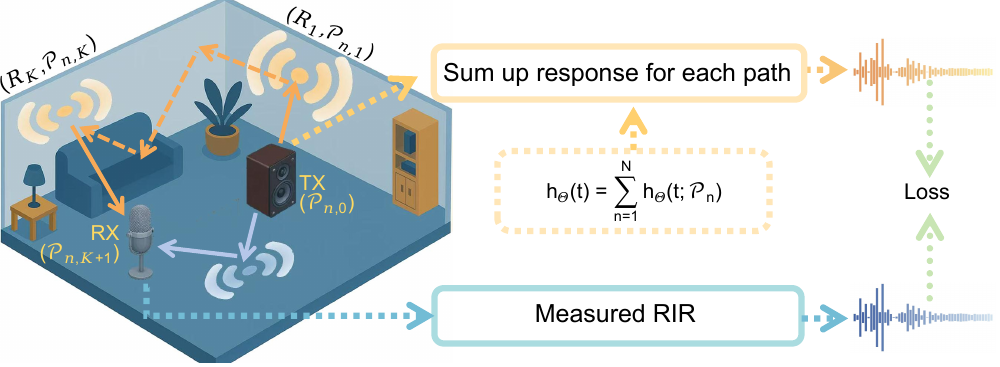}
    \caption{Illustration of differentiable ray tracing for material estimations.
    We cumulate the reflection response for each path and sum up responses from all paths to render the RIR and optimize against the measurements.}
    \label{fig:differentiable_rt}
    \vspace{-10pt}
\end{figure}

\header{Path impact function} $\Gamma(t; \mathcal{P}_n, \Theta)$ represents the impulse response along a single propagation path $\mathcal{P}_n$:
\begin{equation}
\Gamma(t; \mathcal{P}_n, \{R_s\}) = \frac{1}{d_{\mathcal{P}_n}} \, \delta\bigl(t - \frac{d_{\mathcal{P}_n}}{c}\bigr)\ast\mathcal{F}^{-1}\bigl(\prod\limits_{s\in\mathcal{P}_n} R_s(\nu)\bigl).
\end{equation}
The right-hand side of the equation consists of three components.
(1) $\frac{1}{d_{\mathcal{P}_n}}$ models the attenuation due to wave propagation,  where $d_{\mathcal{P}_n}$ is the total distance along $\mathcal{P}_n$. 
(2) The time delay is modeled by the shifted delta function $\delta(t - \frac{d_{\mathcal{P}_n}}{c})$, where $c$ is the speed of sound. 
(3) $\mathcal{F}^{-1}\bigl(\prod\limits_{s\in\mathcal{P}_n} R_s\bigl)$ captures the frequency dependent cumulative attenuation along the path.

\header{RIR from multiple paths.}
The full RIR is the summation of $N$ acoustic paths between speaker and microphone: $h_\Theta\!\left(t\right)=\sum_{n=1}^{N}
h_\Theta\!\left(t;\,\mathcal{P}_n\right)$,
where each path impact function $h\!\left(t;\,\mathcal{P}_n, \Theta\right)$ follows the single-path RIR formulation.

\header{Learning objective.}
Given \(M\) measured RIRs $\{\hat{h}_m(t)\}_{m=1}^{M}$ captured at known devices locations and orientations, we optimize the parameter set $\Theta$ by minimizing the discrepancy (i.e., mean square error) between rendered and measured RIRs.
This objective encourages the renderer to match both the time delays caused by the propagation and amplitudes attenuation observed in the measurements.
The whole process of differentiable ray tracing is illustrated at Fig.~\ref{fig:differentiable_rt}.

\header{Vision priors from visual digital twin.} To inject semantic structure into acoustic reconstruction, we leverage vision priors from the scanned room mesh $G$. 
Rather than treating each small surface patch independently, we leverage surface appearance (i.e., color) consistency to group visually similar and spatially adjacent regions that are likely to share the same acoustic material properties.
To achieve this, we start from seed surfaces with stable colors and normals, we grow regions by iteratively adding neighboring surfaces whose appearance matches the current one~\cite{rabbani2006segmentation}.  
This vision-guided grouping aggregates entire walls, blackboards, wooden surfaces, and other visually coherent objects into unified segments.  
Each segment is assigned with a single reflection parameter.
As shown in the middle column of Fig.~\ref{fig:optimized_material}, our method segment the original mesh into instances that are color-coded differently.

\subsection{Editing Audio-Visual Scene}
\label{sec:audio_visual_scene_editing}

\begin{figure}[t]
    \centering
    \includegraphics[width=0.8\linewidth]{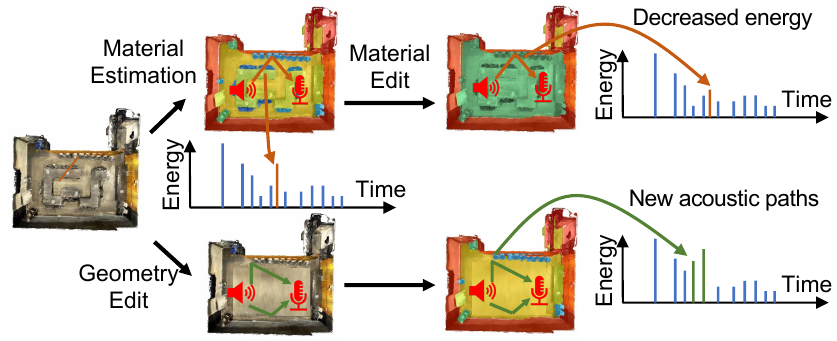}
\caption{Illustration of two types of audio–visual scene edits. Material edits alter the energy of reflected waves without changing their arrival times, whereas geometry edits affect acoustic paths and the resulting RIR.}
    \label{fig:audio_visual_scene_edit}
    \vspace{-15pt}
\end{figure}

Once acoustic properties and scene geometry are reconstructed, \name{} enables editing of the audio–visual scene.
We can supports a wide spectrum of scene manipulations including general mesh deformation, animation, insertion or removal of objects, and modification of acoustic property.
As shown in Fig.~\ref{fig:audio_visual_scene_edit}, we demonstrate two editing capability below and the results in the application (\S\ref{subsec:evaluation_applications}).

\header{Material editing.}
Material editing modifies how scene surfaces interact with sound while keeping the underlying geometry fixed.  
Each mesh segment is assigned a frequency-dependent reflection coefficient $R_s(\nu)$, estimated through the optimization procedure described in \S~\ref{sec:acoustic_property_capture}.  
Users can alter these reflection coefficients of surfaces by reassigning material labels or adjusting to new reflectivity parameters $R_s'(\nu)$.  
For any path $\mathcal{P}_n$, the cumulative attenuation term changes from
$
\prod_{s\in\mathcal{P}_n} R_s(\nu)
\rightarrow
\prod_{s\in\mathcal{P}_n} R_s'(\nu),
$
which alters the amplitude of the RIR while maintaining the propagation delays, as shown in the top row of Fig.~\ref{fig:scene_editing}.
Examples include replacing drywall as absorptive panels, or increase the reflectivity to introduce more reverberations.
After a material edit, the differentiable renderer recomputes the affected propagation paths and synthesizes updated RIRs that reflect the revised acoustic properties.

\header{Geometry editing.}
Geometry edits alter the spatial configuration of the environment and therefore influence the acoustic wave propagation in the environment.  
Supported geometric edits include inserting or removing walls, moving furniture, modifying room layout, or introducing new surfaces such as partitions or acoustic diffusers.
These edits yield updated path sequences $\mathcal{P}_n'$ and path lengths $d_{\mathcal{P}_n}'$, which change both the timing and the ordering of reflections.
Both inserted or removed objects will introduce some new paths while blocking some original paths, resulting in a different RIR (shown in bottom row of Fig.~\ref{fig:scene_editing}.
For example, add furniture in a room will increase the clarity of the sound since there is less echo between wall and floor in the scene.
\vspace{-5pt}
\section{Evaluation}
\label{sec:evaluation}
We first evaluate the performance of audio-visual digital twin including mobile RIR capture (\S\ref{subsec:microbenchmarks}) and acoustic field reconstruction with novel view acoustic synthesis task (\S\ref{subsec:acoustic_field_reconstruction}).
We then evaluate the modifiable audio-visual scene with acoustic property estimation (\S\ref{subsec:evaluation_acoustic_property_estimation}) and demonstrate its editing capability (\S\ref{subsec:evaluation_audio_visual_scene_editing})
We then show applications on immersive auditory experience and acoustic localization (\S\ref{subsec:evaluation_applications}).


\vspace{-5pt}
\subsection{Performance of Mobile RIR capture}
\label{subsec:microbenchmarks}

\begin{figure}[t]
    \centering
    \includegraphics[width=0.7\linewidth]{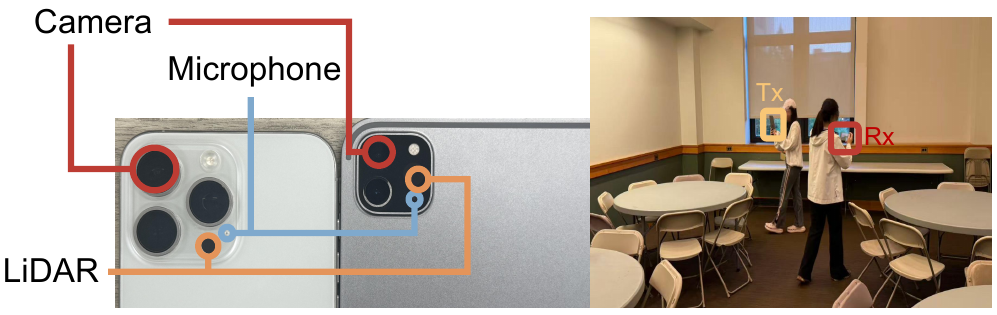}
    \caption{Left: Mobile RIR capture on commodity mobile devices with built-in sensors; Right: Users scan the scene with their phones.}
    \label{fig:hardware_setup}
    \vspace{-10pt}
\end{figure}

\header{Mobile platform implementation.} 
We implement the full acoustic two-way handshake pipeline and visual SLAM within a standalone iOS App that runs on both iPhone Pro and iPad Pro. Our prototype uses an iPhone~15~Pro~Max and an iPad~Pro (4th generation), each equipped with a LiDAR sensor. RTAB-Map is configured to update at 5~Hz to provide accurate real-time pose tracking during mobile scanning.
For audio capture, the device plays chirps through the built-in loudspeaker and records using the rear microphone. 
Chirp is transmitted every 2~s to ensure the previous RIR response has fully decayed before the next excitation.
For the acoustic protocol, the chirp signals are tailored to smartphone hardware limits (<20~kHz) and noise robustness: a high-frequency synchronization chirp (11–19~kHz) for ToF estimation, and a low-frequency chirp (50~Hz–9~kHz) for RIR extraction covering speech and everyday acoustic content. Both chirps last 0.2~s, short enough to avoid motion distortion at typical walking speeds.
Fig.~\ref{fig:hardware_setup} illustrates a typical deployment in which users move through the environment while the App collects synchronized RIRs and trajectories.

\begin{figure}[t]
    \centering
    \includegraphics[width=0.8\linewidth]{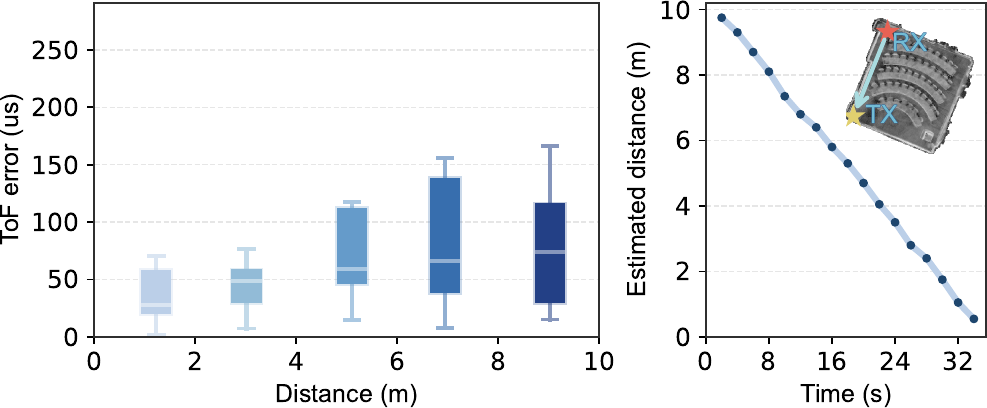}
    \caption{Evaluation on the ToF estimation. Left: we measure the ToF error which has an average of 100~us. Right: a case study where Rx is moving steadily towards the Tx and the estimated ToF decrease gradually.}
    \label{fig:tof_error}
    \vspace{-15pt}
\end{figure}

\header{ToF estimation error.}
We compute the ground-truth ToF by measuring the distance between the Tx and Rx device with a laser meter and dividing it by the speed of sound.
As shown in the left of Fig.~\ref{fig:tof_error}, the estimation error remains stable and does not introduce much growth with increasing distance.
At a separation of 9~m between Tx/Rx, the average ToF error is 100~us, corresponding to 4~cm ranging error given the speed of sound.
On the right of Fig.~\ref{fig:tof_error}, we present a case study where the Rx device steadily moves toward the fixed Tx device.
The estimated ToF and corresponding distance decrease linearly over time, confirming the consistency of measurement under device movement.

\header{RIR collection rate.}
We evaluate the reliability of on-device RIR collection in real environment across varying Tx/Rx distances (1~m to 15~m). 
During a 30-minute scanning session, our system successfully detects the probing chirp $c_1$ with a high detection rate of 99.6\%.

\header{Power consumption.}
We evaluate the power consumption of the App to understand its practicality for everyday use.
With a fully charged iPhone, it can operate for 3 hours of continuous capture.
This corresponds to scanning around 18 rooms, assuming a session length of 10 minutes per room. 

\header{RIR similarity metrics.} 
We quantify the distance between two RIRs with the following objective metrics including energy based reverberation time (T60), clarity (C50) and early decay time (EDT).
We also evaluate the waveform similarty with Envelope (Env) error, FFT Amplitude (Amp) error, Multi-scale STFT (STFT) error that are commonly used in previous work~\cite{lan2024acoustic, luo2022learning, su2022inras}.
Across all these metrics, lower values indicate the two RIRs are similar. We will use these metrics to evaluate whether human users will influence the RIR capture quality and the performance of acoustic field reconstruction.

\begin{figure}[t]
    \centering
    \includegraphics[width=0.8\linewidth]{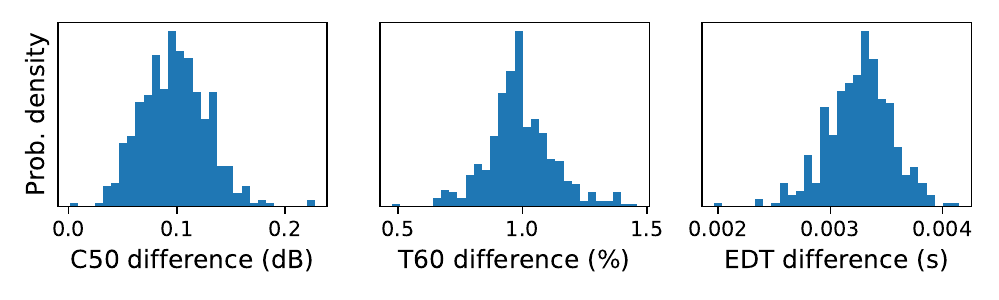}
    \vspace{-5pt}
    \caption{Impact of human presence on measured RIRs. We compare the RIR distance w/o and w/ human user during collection.}
    \vspace{-15pt}
    \label{fig:rir_error}
\end{figure}

\header{Influence of human presence.}
To quantify whether the user’s body affects RIR capture when holding the smartphones, we directly compare handheld measurements against reference tripod measurements across five indoor scenes. As shown in Fig. \ref{fig:rir_error}, the differences in C50, T60, and EDT remain very small and are tightly concentrated around their means: C50 varies by only ~0.1dB, T60 by ~1\%, and EDT by ~3ms. 
This is because the phone’s speaker and microphone are strongly front-facing, so off-axis obstacles (human body) contribute little to the dominant propagation paths. Moreover, at typical listening frequencies, long acoustic wavelengths diffract around the human body, further reducing any measurable influence on the RIR.

\vspace{-5pt}
\subsection{Performance of Acoustic Field}
\label{subsec:acoustic_field_reconstruction}

We evaluate the acoustic field performance via novel view acoustic synthesis. We compare the similarity between rendered RIRs from acoustic field and unseen measured ones.

\begin{figure}[t]
    \centering
    \includegraphics[width=0.85\linewidth]{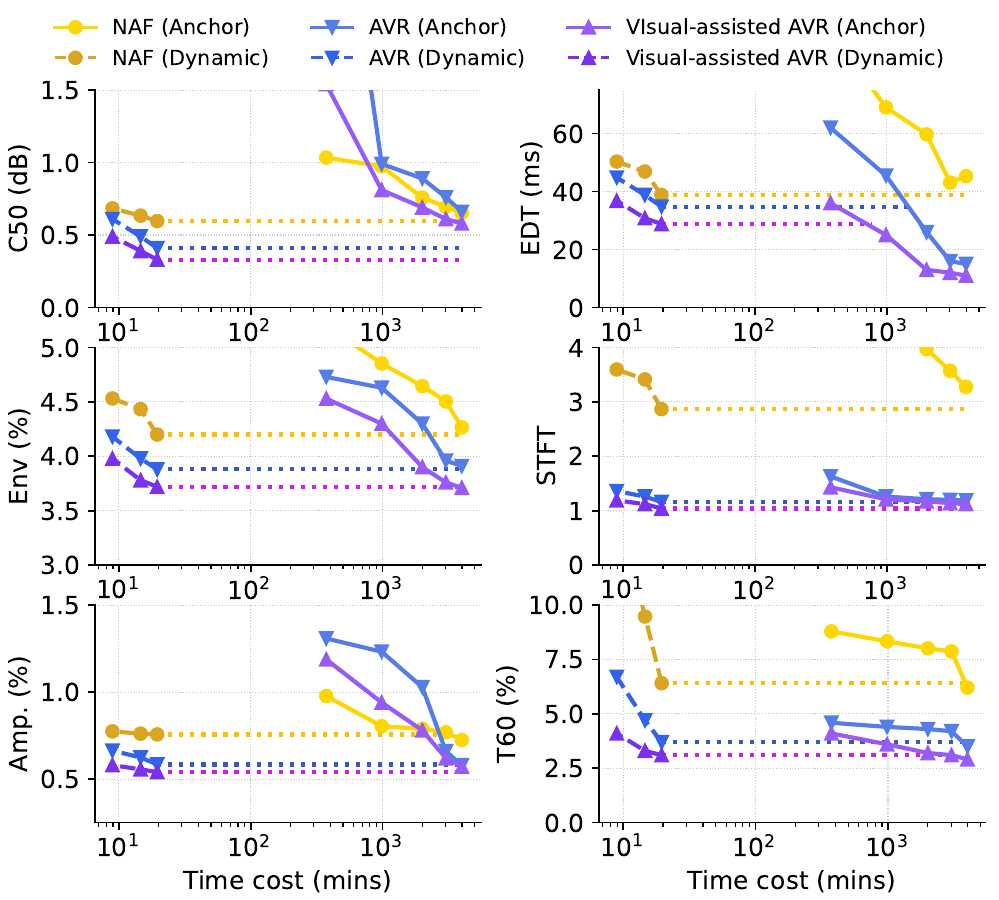}
    \caption{Validation results of different neural acoustic field models trained on 
    dynamic-trajectory datasets versus grid-based datasets. 
    The x-axis denotes the data collection time (log scale), and the y-axis shows reconstruction error across multiple acoustic metrics.}
    \label{fig:acoustic_field_curve}
    \vspace{-10pt}
\end{figure}

\header{Experiment setup.} 
We use two different acoustic capture setup to verify the data efficiency of \name{}.
One is our dynamic trajectory method, where each of the Tx and Rx device is held by a moving user.
The user can freely traverse through the scene while capturing RIRs.
In this setup, the recording session lasts about 20 minutes with a total of 600 samples. 
The other setup is grid-based, consisting of 20 fixed Tx locations and 100–200 Rx positions per location, yielding a total of 2k–3k RIRs.
In this setup, we put the Tx device at a fixed location and let the other user hold Rx device to collect RIRs.
which simulates the traditional RIR measurement setup.
We collect RIR dataset with both setups in four different environments on our campus, containing typical indoor structure such as desks, chair, blackboard, and small objects.
We hold out 10\% of data from both the dynamic-trajectory and grid-based setups to form the test set and train each acoustic model on the two remaining datasets (90\%) separately.
Once the training is done, we let the model to synthesize RIRs at unseen locations to verify the performance.
All the trainings are done with one L40 GPU.

\begin{figure}[t]
\vspace{-5pt}
    \centering
    \includegraphics[width=0.75\linewidth]{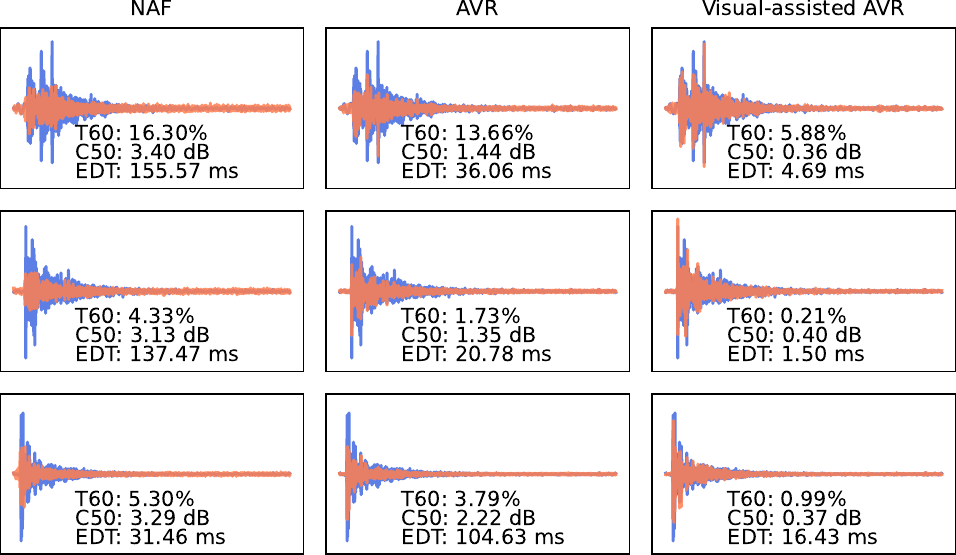}
    \caption{Examples of synthesized impulse response with different methods. Orange is the model predictions, blue is the ground truth ones.}
    \label{fig:rir_visualization}
    \vspace{-15pt}
\end{figure}

\begin{figure}[t]
    \centering
    \includegraphics[width=0.85\linewidth]{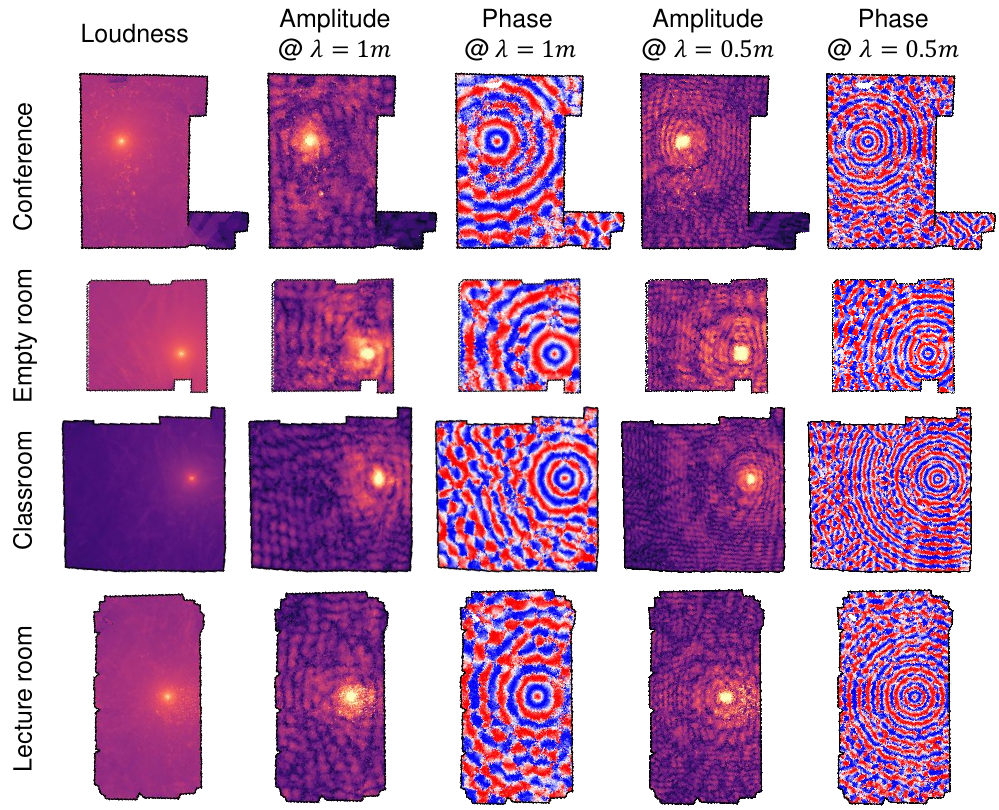}
    \caption{Spatial signal distribution for the estimated acoustic field. We plot the loudness map in the room and the amplitude and phase distribution for the acoustic field at two different wavelengths $\lambda=$1~m and 0.5~m.
    }
    \label{fig:acoustic_field}
    \vspace{-15pt}
\end{figure}

\header{Results.}
Fig.~\ref{fig:acoustic_field_curve} presents the performance of three neural acoustic field models (NAF, AVR and visual-assisted AVR) when trained on datasets collected via our dynamic-trajectory method versus the traditional method. 
The x-axis denotes the time required for dataset collection, while the y-axis shows error across multiple metrics.
Results consistently demonstrate that dynamic-trajectory method are far more data-efficient. 
For the same acoustic model, training on our 20-minute dynamic-trajectory dataset often achieves comparable or superior performance to training on grid-based datasets collected over 3000 minutes.
This corresponds to more than 100× reduction in collection time.
This finding confirms that continuous motion supplies more diverse RIR measurements compared to grid-based method.
Furthermore, training visual-assisted AVR with 10~mins data can almost achieve similar performance of AVR when training on 20~mins, further improving the data efficiency.
Fig.~\ref{fig:rir_visualization} shows the visualization of rendered RIRs for each method. 
Visual-assisted AVR can render RIR with better alignment to the ground truth ones.
Besides, we also compare the rendering speed between AVR and visual-assisted AVR on a L40 GPU.
While AVR takes 100~ms to render a single RIR, our method only takes 10~ms to output the same RIR, improving the rendering efficiency by 10x.
We also show examples of learned acoustic field in Fig.~\ref{fig:acoustic_field}, where we plot the loudness map, amplitude and phase map at two different wavelengths, which shows complex wave propagation phenomenon.

\subsection{Performance of Property Estimation}
\label{subsec:evaluation_acoustic_property_estimation}

\begin{figure}[t]
    \centering
    \includegraphics[width=0.9\linewidth]{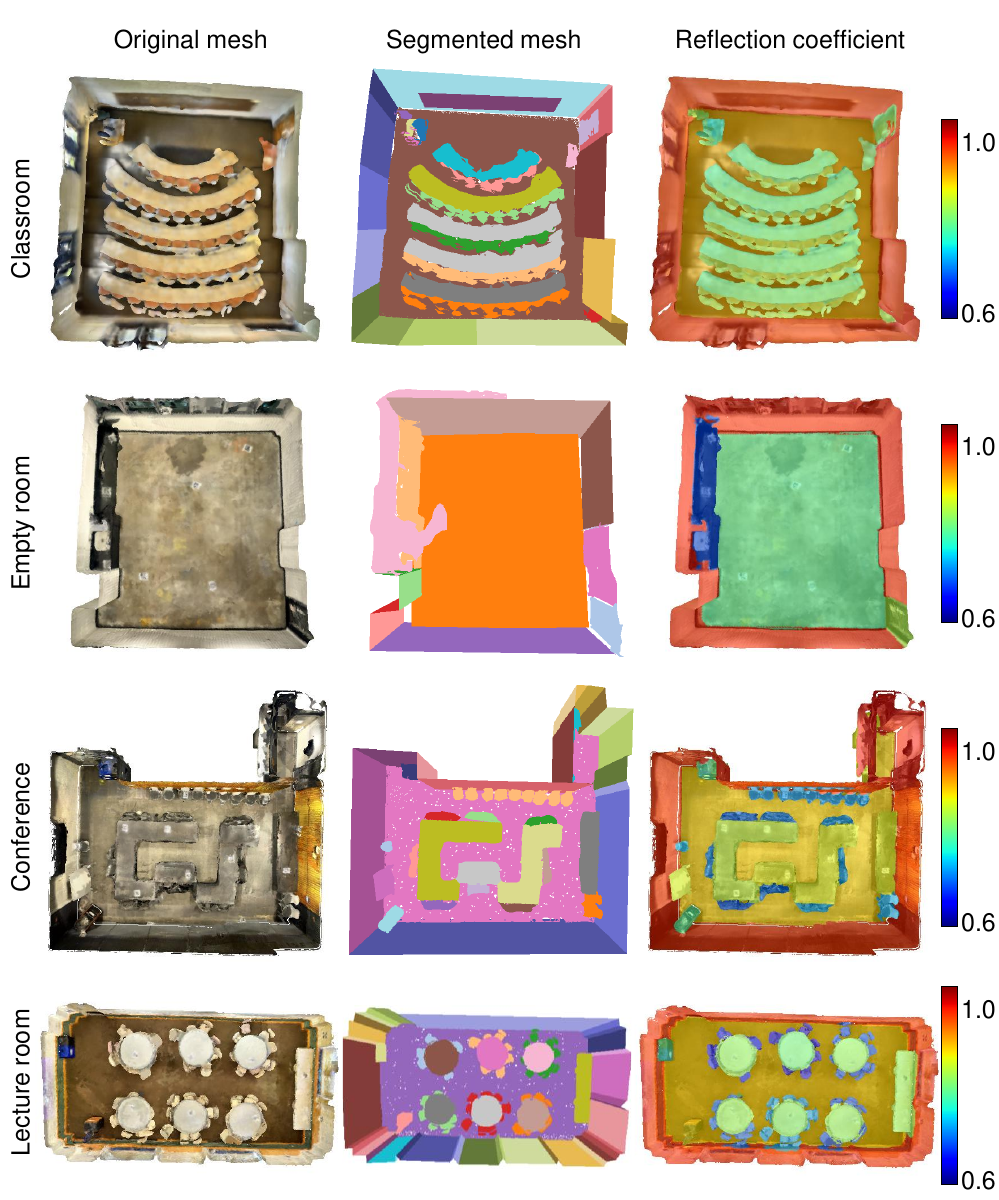}
    \caption{Visualization of material estimations. We show top-down view for each scene at different rows. 
    First col: original mesh; Second col: segmented instances are color coded separately;
    Third col: estimated reflection coefficients, blue and red indicate low and high reflectivity, respectively.}
    \label{fig:optimized_material}
    \vspace{-10pt}
\end{figure}

\header{Experiment setup.} 
We build differentiable acoustic rendering based on AcoustiX simulation~\cite{lan2024acoustic}.
We enumerate specular reflections up to eight bounces.
Each segmented surface is associated with a sets of learnable reflection coefficient $Rs$ at frequency of 125, 250, 500, 1k, 2k, 4k, 8k. The rest frequency response are linearly interpolated.
Tx/Rx pattern is parameterized by a set of spherical harmonics.
We evaluate this framework on four indoor environments.
The geometry and speaker/microphone positions are fixed and not subject to optimization.
We use a RTX 4070 GPU for optimizations.

\begin{figure}[t]
  \centering
  \small
    \begin{minipage}[c]{0.55\linewidth}
      \centering
      \includegraphics[width=0.8\linewidth]{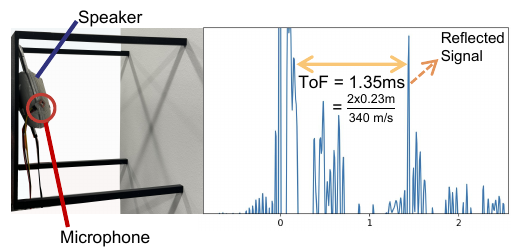}
      \caption{Experiment setup to get the ground truth reference measurement.}
      \label{fig:material_setup}
    \end{minipage}\hfill
    \begin{minipage}[c]{0.43\linewidth}
      \centering
      \captionof{table}{Results on material reflection coefficients estimations.}
      \begin{adjustbox}{max width=0.8\linewidth}
        \begin{tabular}{ccc}
          \toprule
          Material & Fixed setup & Estimated \\ \midrule
          Drywall & 0.90 & 0.93 \\
          Glass   & 0.88 & 0.92 \\
          Metal   & 1.00 & 0.94 \\
          Wood    & 0.81 & 0.88 \\
          Carpet  & 0.71 & 0.85 \\ \midrule
        \end{tabular}
      \end{adjustbox}
      \label{tab:material_estimations_results}
    \end{minipage}
  \vspace{-15pt}
\end{figure}

\header{Ground truth measurement.}
To obtain reference reflections for evaluation, we perform controlled measurements using co-located Tx/Rx pair positioned 23\,cm from each surface.  
A broad band chirp signal is emitted toward the surface, and the reflected signal is recorded.  
From each recording, we isolate the reflection and normalize it by a constant factor to produce a relative reflection measurement.  
Though this value is not an absolute reflection coefficient, it provides a consistent reference across surfaces.  
We can use this measurement to assess whether our estimated coefficients are linearly correlated with the fixed setup. We show the setup and the process to get the second reflection at Fig.~\ref{fig:material_setup}.

\begin{figure}[t]
    \centering
    \includegraphics[width=0.85\linewidth]{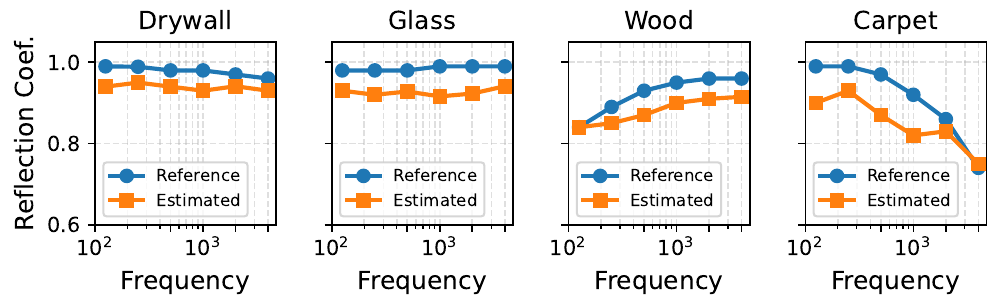}
    \caption{Comparison of estimated reflection coefficient and reference one for each material across different frequency.}
    \vspace{-10pt}
    \label{fig:reflection_coefficient}
\end{figure}

\header{Results.} 
Fig.~\ref{fig:optimized_material} shows the textured mesh, segmented mesh and the estimated reflection coefficient (averaged across all frequency).
Across all four rooms, the estimated reflectivity aligns with prior reports.
Enclosure walls (drywall) are consistently the most reflective; floors with carpeting appear are substantially less reflective than surrounding walls; and movable furniture, such as tables and chairs that are made of wood, tend to exhibit lower reflectivity.
Fig.~\ref{fig:reflection_coefficient} shows the estimated reflection coefficient compared with the reference one~\cite{standard1990standard, jcw_absorption} across four common materials. 
Results show that the estimated coefficients matched very well with the ground truth one, with a mean absolute error of 5.3\%.
Tab.~\ref{tab:material_estimations_results} quantifies the estimates (averaged across different frequency) and the fixed setup measurement results.
The fixed-setup measurements and our estimated ones exhibit a strong Pearson correlation coefficient ($r\!=\!0.96$).
These results indicate that our method reliably recovers material-dependent reflectivity.

\header{Evaluation on novel-view acoustic synthesis.}
Beyond estimating material properties, our differentiable acoustic renderer is capable of synthesizing high quality RIRs at new position, similar to acoustic field model.
To assess the quality of these rendered RIRs, we compare it against visual-assisted AVR.
Across all evaluation scenes, the two methods achieve highly comparable acoustic metrics: RT60 of 3.3\% (visual-assisted AVR) vs 3.5\%, C50 of 0.32~dB vs. 0.35~dB, and EDT of 33.5~ms vs.35.9~ms.
These results show that the renderer reliably produces realistic RIRs, enabling physically consistent, editable audio–visual scenes.

\vspace{-5pt}
\subsection{Dynamic Audio-Visual Scene Editing}
\label{subsec:evaluation_audio_visual_scene_editing}

\begin{figure}[t]
    \centering
    \includegraphics[width=0.80\linewidth]{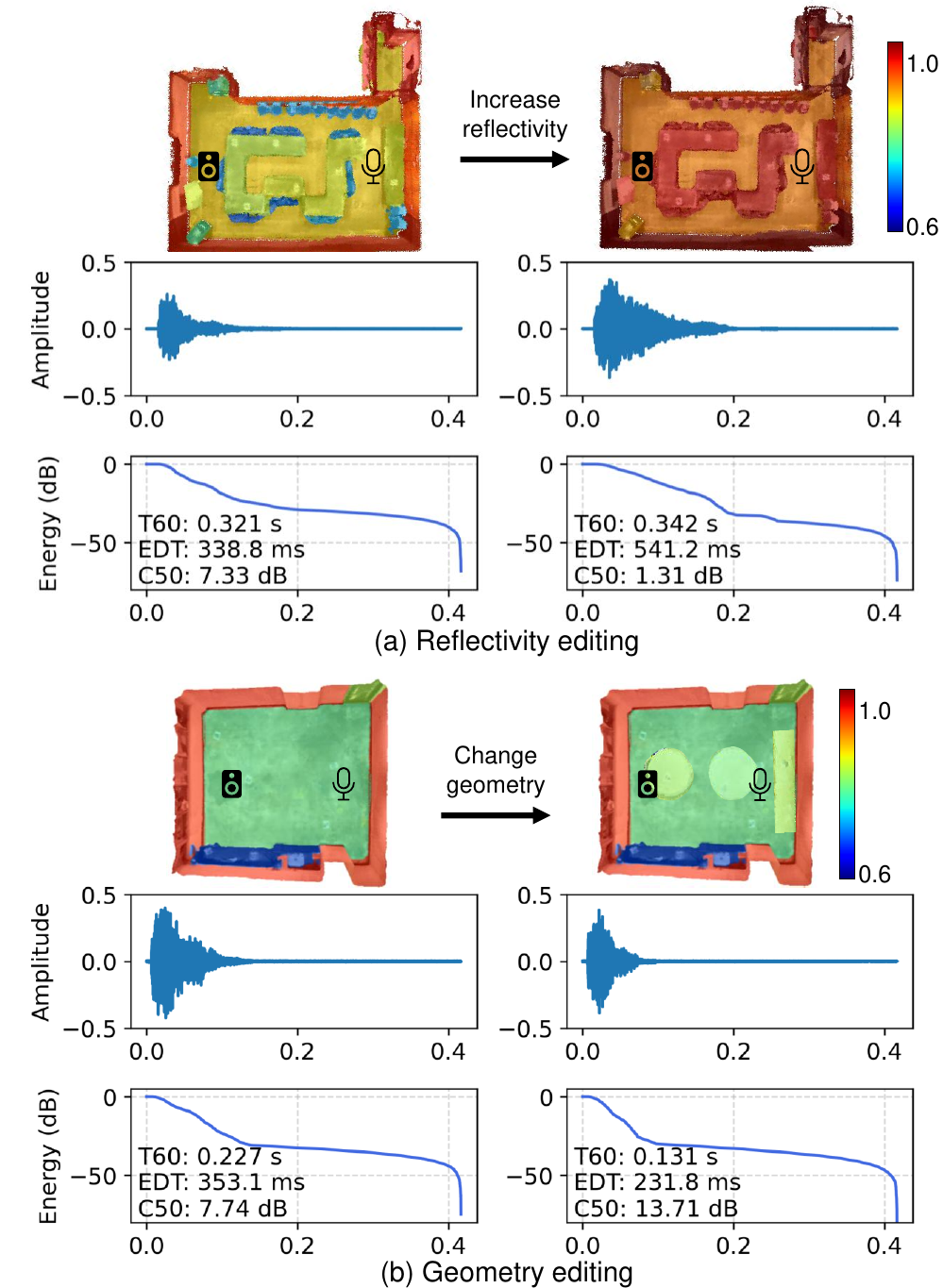}
    \caption{(a) Increasing the reflectivity can increase the reverberation time. (b) Adding extra furniture to the empty room can reduce echoes.}
    \label{fig:scene_editing}
    \vspace{-13pt}
\end{figure}

Our framework enables interactive editing of the audio-visual scene and we provide two cases for this capability. 
One is material property editing, as shown in Fig.~\ref{fig:scene_editing}(a). 
We fix the Tx/Rx positions and increase the reflectivity of all the surfaces in the environment. 
As a result, the rendered RIR exhibits a longer late tail and the energy decay curve for the RIR flattens.
This indicates a space with stronger reverberation, as reflected by the increased RT60 and EDT values and the decreased C50, since early energy constitutes a smaller fraction of the total after editing.
We show another example of geometry editing in the Fig.~\ref{fig:scene_editing}(b), where several tables are added into the empty room.
The added geometry reduces the number of path between walls and floor and introduces more absorptions.
The re-rendered RIR shows a shorter late tail and more rapidly diminishing reflections and the energy-decay curve also becomes steeper.
Perceptually, the room would sound "tighter" and less echoic.
This aligns with common observation where one can hear more reverberations when clap hand loudly in an empty or unfurnished room.
But once the room is furnished, the echo becomes less obvious.
We provide a user study in the next section about audio rendering to assess whether the acoustic editing matches with the visual aspect.

\vspace{-5pt}
\subsection{Applications}
\label{subsec:evaluation_applications}

\subsubsection{Immersive Audio Rendering}\mbox{}

\noindent
One important application for acoustic field capture is to synthesize the immersive audio contents.
Once the acoustic field of a room is captured, we can freely render how a listener would perceive sound while moving through the environment. 
Along any given trajectory the user would experience in the scene, we first associate each receiver pose with its corresponding RIR by querying the acoustic field model.
A dry source signal (e.g., speech or music) is then convolved with each RIR to generate short audio segments that capture the acoustic effects at that location. 
At the same time, the trajectory can be used to render synchronized visual frames from the reconstructed mesh, so that both sound and visuals evolve consistently as the user moves through the space. 
This joint audio-visual synthesis provides an immersive reproduction of the scene from arbitrary paths. 
We evaluate the applications on the immersive auditory experience with a perceptual user study.

\begin{figure}[t]
    \centering
    \includegraphics[width=0.8\linewidth]{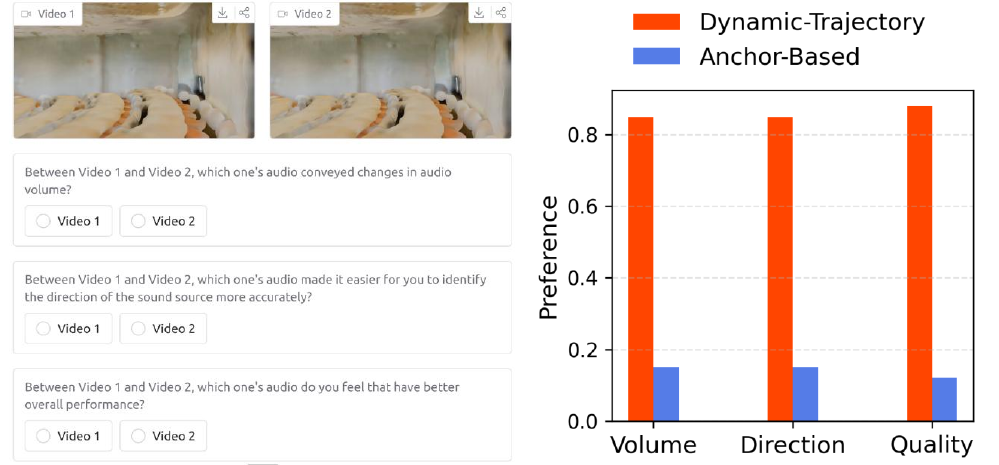}
    \caption{User study to compare dynamic-trajectory and grid-based method. Left: the interface for user study: each user is asked to watch and listen two video pair and answer the questions to compare them; Right: average user preferences from three pair of videos.}
    \label{fig:user_study}
    \vspace{-12pt}
\end{figure}

\header{Setup.} 
To compare dynamic trajectory method and grid-based method, we produce RIRs from acoustic model that are trained on these two datasets.
Each RIR is convolved with a dry sound track to obtain a wet rendered sound, paired with a time-synchronized video captured from the receiver's viewpoint.
Participants $(N=17)$ watch the video with headphones and are then asked to compare the clip along three perceptual dimensions: volume consistency, directional cues, and overall quality.

\header{Results.} $88\%$ of responses preferred dynamic-trajectory over the grid-based method in terms of overall quality. 
$85\%$ of responses indicate that our method outperforms traditional method in sound volume and directional cues.

\subsubsection{Perceptual Evaluation of Audio-Visual Scene Editing}\mbox{}
\label{subsec:evaluation_dynamic_scene_editing}
Beyond demonstrating that our renderer faithfully reflects material and geometry edits in the RIR domain, we further conduct a perceptual user study to assess whether these edits produce intuitive and expected changes in the resulting audio.
We test (i) whether increasing or decreasing global reflectivity produces the expected differences in reverberation, and (ii) whether adding or removing large objects produces perceivable damping or enrichment of room acoustics.

\begin{figure}
    \centering
    \includegraphics[width=0.8\linewidth]{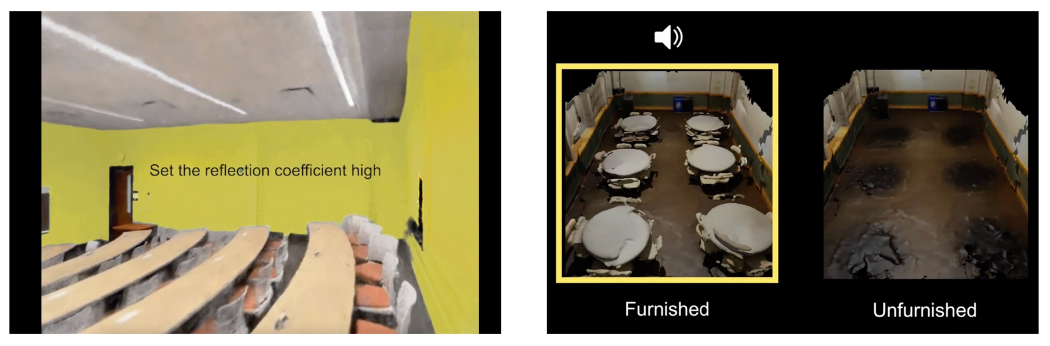}
    \caption{Snapshot of audio-visual scene edit for user study. Left: change the reflectivities of the wall. Right: change the furniture in the room.}
    \label{fig:scene_edit_snapshot}
    \vspace{-15pt}
\end{figure}

\header{Setup.}
For each environment, we define a short camera trajectory and render synchronized audio–visual clips before and after each scene edit.
For material editing, we uniformly increase or decrease all wall reflectivities.
For geometry editing, we add or remove tables from the room mesh.
After edits, we re-run the differentiable rendering pipeline to get the edited RIRs and we follow same procedure to render wet sound in previous section.
Participants ($N$ = 15) wear headphones and watch paired video clips.
For each pair, they answer which clip better matches the visual scene along two dimensions:
(1) perceived reverberation level (e.g., “more echoic”, “more damped”), and
(2) audio–visual consistency (“Does this audio match what you expect from the depicted room?”).
A snapshot of these two edits is shown in Fig.~\ref{fig:scene_edit_snapshot}.

\header{Results.}
For material editing, 93\% of participant responses correctly identified the higher-reflectivity scene as producing more reverberant audio, and 89\% judged the lower-reflectivity edit as sounding more damped.
For geometry editing, 91\% of users feels that adding tables in the room increase the clarity of the sound and unfurnished room sounds more reverberant.
Users all feel that the the edited acoustic rendering match with the visual scene and edits, demonstrating that \name{} produces perceptually coherent audio–visual edits.

\subsubsection{Acoustic Localization}\mbox{}
RIRs inherently encode rich spatial cues like multipath structures, which enable estimating the microphone position from a single Tx/Rx pair.
We demonstrate how acoustic field can  enhance localization.


\header{Acoustic field data augmentation.}
Once we reconstructed the acoustic field, it can synthesize additional RIRs at arbitrary receiver positions.
It can provide denser spatial coverage beyond what is feasible to measure. 
By augmenting the dataset, the localization model is exposed to more diverse multipath patterns and performs much better.

\header{Setup.} We collect data in three environments: a classroom ($8m\times10m$), a lecture hall ($7m\times12m$) and an L-shaped room ($8m\times9m$), each furnished with tables, chairs, etc. 
In each room, we fix the Tx at one location and move the Rx throughout the room for 30 minutes, producing 900 RIRs per room. 
We train the localization model with two variants: (1) trained with raw RIRs and (2) additional RIRs augmented from acoustic field.
We adopt a lightweight 1D CNN that maps a raw RIR waveform to a probability distribution of the microphone location.
We split first 80\% of the data for training and rest for testing. 
For the variant with the field augmentation, we use the same training set to first reconstruct the acoustic field and then synthesize additional 8k RIRs to augment the training.
We evaluate performance in absolute error between predicted and ground-truth positions.

\begin{figure}[t]
    \centering
    \includegraphics[width=0.9\linewidth]{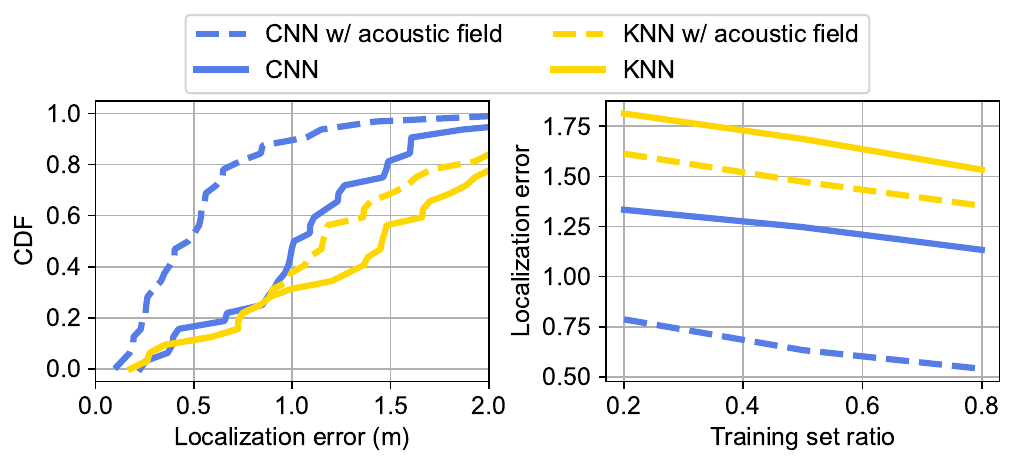}
    \caption{Acoustic localization results. Left: CDFs of localization error. Right: localization error with different training set ratio.}
    \label{fig:localization_result}
    \vspace{-15pt}
\end{figure}

\header{Results.} 
1D CNN-based model can achieve a medium error of 1~m averaged across three rooms, surpassing KNN baseline by about 0.5~m (Left of Fig.~\ref{fig:localization_result}). 
Acoustic field model can further boost their performance and reduce the localization error to 45~cm for CNN-based model.
We also ablate the performance with different training set ratio (right of Fig.~\ref{fig:localization_result}) , acoustic field augmentation can boost the performance by a large margin at various ratio.


\section{Discussions}
\label{sec:discussion}
\header{Limitations.}
Our acoustic field reconstruction and material-parameter estimation currently run offline on a desktop GPU. 
A potential next step is to develop lightweight models that enable on-device training and inference. 
Another promising direction is to infer the full acoustic field directly from one or a few scene images. Achieving this will require data-driven models trained on large-scale audio–visual datasets, which we view as a natural next step enabled by our system.

\header{Conclusion.} In this work, we introduce \name{}, the first practical system for constructing modifiable, audio-visual digital twins using only commodity smartphones. 
\name{} aims to bridge the gap between acoustic twin and visual twin by incorporating visual priors into the acoustic capturing, reconstruction, and editing process. 
\name{} narrows the gap between research setups and everyday devices, and supports various applications that open up a range of opportunities for immersive AR/VR experience, smart homes that were previously restricted to costly professional hardware.


\bibliographystyle{ACM-Reference-Format}
\bibliography{references}


\end{document}